\documentclass[final]{article}
\usepackage{authblk,amsmath,graphicx,booktabs}

\title{The Motivated Can Encrypt (Even with PGP)}

\author[1]{Glencora Borradaile\thanks{glencora@oregonstate.edu (corresponding author)}}
\author[2]{Kelsy Kretschmer\thanks{kelsy.kretschmer@oregonstate.edu}}
\author[1]{Michele Gretes\thanks{mgretes@riseup.net}}
\author[1]{Alexandria LeClerc\thanks{leclerca@oregonstate.edu}}
\affil[1]{School of Electrical Engineering and Computer Science}
\affil[2]{School of Public Policy}
\affil[ ]{Oregon State University}

\date{2021}
\begin{document}
\maketitle

\begin{abstract}
  Existing end-to-end-encrypted (E2EE) email systems, mainly PGP, have long been evaluated in controlled lab settings.
  While these studies have exposed usability obstacles for the average user and offer design improvements, there exist users with an immediate need for private communication, who must cope with  existing software and its limitations.
  We seek to understand whether individuals motivated by concrete privacy threats, such as those vulnerable to state surveillance, can overcome usability issues to adopt complex E2EE tools for long-term use.
  We surveyed regional activists, as surveillance of social movements is well-documented.
  Our study group includes individuals from 9 social movement groups in the US who had elected to participate in a workshop on using Thunderbird+Enigmail for email encryption.
  These workshops tool place prior to mid-2017, via a partnership with a non-profit which supports social movement groups.
  Six to 40 months after their PGP email encryption training, more than half of the study participants were continuing to use PGP email encryption despite intervening widespread deployment of simple E2EE messaging apps such as Signal.
  We study the interplay of usability with social factors such as motivation and the risks that individuals undertake through their activism.
  We find that while usability is an important factor, it is not enough to explain long term use.
  For example, we find that riskiness of one's activism is negatively correlated with long-term PGP use.
  This study represents the first long-term study, and the first in-the-wild study, of PGP email encryption adoption.
\end{abstract}

\section{Introduction}
Controlled lab studies have identified usability issues that are particular to secure communications technology, and addressing these issues can increase security through correct tool use.
Whitten and Tygar's seminal ``Johnny Can't Encrypt'' paper, as well as replications of their method, have found persistent usability issues with end-to-end encrypted (E2EE) PGP email over the past 20 years \cite{whitten_why_1999, sheng_why_2006,  ruoti_why_2015, garfinkel_johnny_2005}.
Others have sought to understand what roles key exchange, trust, and transparency have on the usability of PGP~\cite{bai_inconvenient_2016, ruoti_we_2016, atwater_leading_2015}.
Each of these studies were in a time-limited laboratory setting with participants who are largely first-time users recruited from university campuses.
These studies tend to be forward-looking through suggesting design improvements and testing proof-of-concept prototypes.
However, despite usability issues, there exists a population of users with an immediate need for private email communication who cannot wait for improvements to be made to existing PGP email solutions.
Activists, journalists and lawyers have a need to use secure communication technology, specifically email, but often face obstacles in adoption~\cite{lerner_confidante_2017}.
E2EE messengers such as Signal \cite{greenberg_signal_2015} provide easier-to-use alternatives, but because these tools lack certain functionality, users continue to opt for less secure alternatives in order to work effectively~\cite{luca_expert_2016, bai_inconvenient_2016, mcgregor_investigating_2015, marczak_social_2017, abu-salma_exploring_2018}.

Surveillance against social movement work has been long-standing and persistent~\cite{cunningham_there_2004, marczak_when_2014, marczak_hacking_2015, marczak_social_2017, deibert_communities_2014, scott-railton_security_2016}.  The US government has long been known to use surveillance to support interference of social movements as was notable during the Civil Rights Era through COINTELPRO, resulting in such extreme outcomes as the murder of Black Panther Party leaders.  Surveillance and interference tactics have continued into the modern era~\cite{potter_green_2011}, despite formal protections through the First and Fourth Amendment that should protect dissent and public engagement through activism.  Only E2EE provides mathematically formal protections against illegal search.

Starting in 2014, in partnership with the Civil Liberties Defense Center (CLDC), a non-profit which provides legal support to US social movement groups, a subset of our research team developed and deployed digital security workshops for social movement groups including one on PGP email encryption. The CLDC encourages the use of E2EE technologies as robust backup to the legal protections of ``I do not consent to this search'' and to counter the chilling effect created by mass surveillance~\cite{penney_chilling_2016}.
Our experience with the CLDC's use of PGP and with workshop participants' use of PGP over the span of several years suggested that these users had better success with PGP email than PGP usability research would indicate.
We deployed a follow-up survey of PGP workshop attendees in order to understand if this was true, and if so, why.

In this research, we ask: How do users with concrete privacy threats respond in the long term to training that aims to overcome the documented usability issues of PGP email encryption?

We do not seek to uncover what aspects of a particular tool should be changed to increase the ease with which a user can learn to use the tool for the first time as in PGP usability studies. Rather we seek to understand what conditions lead to success or failure to overcome usability issues with PGP email encryption in the long term.  We hypothesize that our workshop participants are more motivated to get PGP working than the average user and that this motivation stems from the activities in which they engage.

We adopt a method of analysis from political science to understand the combinations of conditions that explain the outcome (adopting or not adopting long-term PGP use).  While we consider ease of use as one factor, we explore the interplay of this factor with other factors such as motivation and the risks that individuals are undertaking (Section~\ref{sec:fact}).  We look at two types of motivation: individual (wanting to protect one's own information) and group (wanting to protect a group's information or activities). We look at risk in terms of the potential consequences (including loss of employment, fines or criminal charges).  We describe how we measure these factors in Section~\ref{sec:measure}.

We perform a Qualitative Comparative Analysis (QCA, Section~\ref{sec:qca}).  QCA is not a statistical method, but is rather a reproducible means of organizing descriptive data for qualitative analysis. QCA is appropriate for understanding a small number of detailed cases, such as our sample ($n=19$).

We find that the participants who adopted PGP in the long term are characterized by finding PGP relatively easy to use, {\em and}
\begin{itemize}
  \item being individually motivated to adopt PGP, {\em or}
  \item not
    engaging in activism or engaging only in low-risk activism.
  \end{itemize}
  On the other hand, the participants who did {\em not} adopt PGP in the long term are characterized by:
  \begin{itemize}
  \item finding PGP difficult to use {\em and} belonging to a community using other secure communications technology; {\em or}
  \item finding PGP difficult to use {\em and} engaging in high-risk activism; {\em or}
  \item not being individually motivated to adopt PGP {\em and} engaging in high-risk activism.
\end{itemize}
The negative correlation with risk is disappointing, as digital security trainers would prefer that higher-risk individuals take their digital security more seriously (see Section~\ref{sec:risk}).

This analysis uncovers aspects that may help guide digital security trainers: certainly there is a need to overcome usability barriers, but it may also help to understand what brought an individual to the training and tailor the training to that.  For example, making sure high-risk individuals understand what they may individually gain through increasing their digital security.

Our study is the first long-term study of PGP email encryption.  It is also the first in-the-wild (e.g.\ out of the laboratory) study of PGP email encryption adoption as our participants were not research subjects at the time of training.  Our study population is unique, and a challenging one from which to recruit. We recruited 17 members of social movement organizations, including direct-action groups, and 2 individuals not engaged in social movement organizing but who participated in the workshops.
These subjects provide insight into the ability of motivated individuals to overcome usability issues in encryption technologies. This work also reflects the importance of motivation, as found from McGregor et al.~\cite{mcgregor_when_2017} as well as contextualizes the obstacles outlined by Renaud et al.~\cite{renaud_why_2014} and Abu-Salma et al.~\cite{abu-salma_obstacles_2017} in the uptake of secure communication.

\section{Context and Motivation}\label{sec:setting}

The focus of this work is activists who use email for their organizing and whether they adopt PGP email encryption in the long term following a training.  In this section we give context for why this is our focus and why it is relevant.  Partly, the focus on email and PGP is a result of when this research started (before simpler E2EE communication options were widely used).  However, despite the movement of activist organizing onto E2EE messaging apps, many activist groups continue to use email in addition to other targeted groups (such as journalists and lawyers).

\subsection{Why activists?}

In social movement groups across the political spectrum, individual activists frequently valued both transparency in their organizing work, and have deep concerns about surveillance and privacy.
For example, Agarwal et al.~\cite{agarwal_grassroots_2014} found that members from Tea-Party groups (on the political right) and communities spurring from Occupy Wall Street (on the political left) expressed similar concerns about security of both group and personal information, but at the same time they wanted their group's decision-making processes to be transparent.
This creates a critical dilemma for activists; they need open communication within their communities, at the same time those communities are increasingly fearful of surveillance from corporations and the government.
Members must find a balance for these conflicting needs given the risk that members will disengage if they fear their information is not secure, or when they feel their fears are not attended to by leaders~\cite{rohlinger_fervor_2016, agarwal_grassroots_2014, mcgregor_individual_2016}.

These fears are not unfounded. Digital attacks on social movement groups have been well documented \cite{deibert_communities_2014, scott-railton_security_2016, marczak_when_2014, marczak_social_2017, marczak_hacking_2015}, and digital surveillance has been able to breach what were previously established safe-havens from surveillance~\cite{marczak_hacking_2015}. The integration of technology into everyday life allows for greater, and richer data collection, that requires more expertise to secure.

The unique demographics and needs of activists in terms of socio-economic status, education, and access to technology influence how activists use technology. In her study of how class shapes digital activism,  Schradie~\cite{schradie_digital_2018} found that working class activists frequently lacked the financial means to develop internet communication skills and, when faced with technological challenges, lacked the knowledge to fix them \cite{ortiz_giving_2019, george_clicktivism_2019}. Self-education can also be complicated; even though information on secure practices is widely available, users have a hard time prioritizing what is important, and what they do prioritize is often not aligned with what experts recommend \cite{redmiles_comprehensive_2020}.
Additionally, even established organizations that support or conduct social movement work often face the same threats as private sector or government organizations, but have access to far fewer resources by comparison \cite{deibert_communities_2014, scott-railton_security_2016}.

Compared to the participants of evaluations held on university campuses, users involved in social movements differ from what is considered the average user. Their risk level, including potential consequences may provide stronger motivation for secure communication than the typical user.  While this may mean our observations would not generalize to {\em average} users, we believe that our observations will have impacts for at-risk groups (e.g. lawyers, journalists and broader communities of resistance), which we discuss further in Section~\ref{sec:discuss}.

\subsection{Why email?}
In most groups, email serves as a vital tool in their organizing work and communicating openly with other community members.
Activists find email useful in every part of social movement activism, from debating issues within groups, announcing actions, and planning for the future.
Research in both left-wing and right-wing activist groups have found that activists prefer to communicate by email because it is widely available and offers the potential for higher levels of privacy, especially in comparison to carrying out these organizing tasks on social media sites~\cite{agarwal_grassroots_2014}.
Despite its primacy among activists, many remain concerned about the threat of government and corporate surveillance of email.  At the time of our PGP email workshops (starting in 2014) from where we recruit participants, the simple plug-and-play E2EE apps that exist now were not available as potential alternatives to insecure email, and group messengers such as Slack were not yet in popular use.  Even though simpler options were available at the time of the survey (in 2018), the groups supported by our collaborating non-profit were still using email for organizing.

\subsection{Why PGP?}

End-to-end encrypted email, in the form of PGP email, offers protection against this surveillance, and applications implementing PGP have been available for years.
However, the seminal paper by Whitten and Tygar \cite{whitten_why_1999} and its replications, \cite{ruoti_why_2015, sheng_why_2006, garfinkel_johnny_2005} have consistently found PGP email difficult to uptake for the average user. While originally theorized to be limitations of the user interface, other hypotheses have been examined as to why PGP is difficult to uptake such as: the complexity of key management \cite{bai_inconvenient_2016, garfinkel_johnny_2005}, the reliance on popular email platforms \cite{atwater_leading_2015, ruoti_we_2016}, as well as how the system displays feedback to the user \cite{ruoti_confused_2013, bai_inconvenient_2016}.  As of our PGP email workshops, there were not any readily available solutions to these established problems.

Out of necessity to continue to use email, the CLDC promoted PGP email encryption for their supporting groups, given their own internal success with its use.  While general-use PGP has been studied from a usable-security perspective as just highlighted, research on PGP email focused toward targeted groups such as social movements is lacking, with a few notable exceptions:  Gaw et al.\ studied a single non-violent, direct-action organization ethnographically~\cite{gaw_secrecy_2006}; Lerner et al. sampled lawyers and journalists in order to evaluate a prototype PGP email client~\cite{lerner_confidante_2017}; and, McGregor et al. investigated the success of security measures, including PGP, in the Panama Papers project~\cite{mcgregor_when_2017}.  We discuss how our work relates to these three papers in the Section~\ref{sec:discuss}.

\section{Hypotheses}\label{sec:fact}
Our study seeks to understand the real communication choices of people operating in particular social circumstances and provides deeper insight into how individual and collective conditions jointly shape technology decisions.
In this section, we discuss a mix of individual and collective conditions that are likely to affect technology decisions and which we study in this work.  We examine their relationship to social movement participants in particular and hypothesize how these conditions would impact the adoption of PGP email encryption.

\paragraph*{Ease of Adoption}
Given the well-documented difficulty with learnability of PGP email encryption, we expect that an individual's {\bf ease in adopting new technologies} and {\bf technological confidence} should be the most important condition determining continued use.
In their work on the role of technology among activists, Dencik and Cable~\cite{dencik_advent_2017} found that while individual group members were concerned about surveillance, they frequently chose to coordinate their activism on widely available and insecure social media platforms, reporting that their lack of technical ability meant reliance on widely available and user-friendly platforms that were known to be insecure.
Activists' lack of access to resources (including  technology and IT support) can also result in lower tech ease and confidence~\cite{schradie_digital_2018, ortiz_giving_2019, george_clicktivism_2019}.
Given the documented gaps in technical skills and abilities among activists,
{\em we expect that the ability to adopt the tool easily to be important for continuing to use encryption technologies.}

\paragraph*{Personal Motivation}
Beyond technical abilities, people vary significantly in their \textbf{personal motivation} to protect their communications from surveillance \cite{dencik_advent_2017,dencik_data_2016}.
Personal motivations matter a great deal in social movement contexts.
Activists often work in groups, and the security practices of the collective are only as strong as the commitment of all individual members \cite{dencik_data_2016,leistert_resistance_2012}.
{\em We expect that the presence of personal concern or fears about surveillance to be important for adopting and continuing to use encryption technologies.}

\paragraph*{Group Motivation}
Just as personal motivation varies across individuals, we should expect that some respondents will be more concerned about protecting their group's communication than their personal communications.
Individuals who are cognizant of the communal benefits of encryption (offering protection from surveillance for whole communities)---that is, individuals with \textbf{group motivation}---are likely to find their social relationships and their role in their communities as salient motivations for pressing through difficulties with encryption technologies.
Previous studies have revealed that a strong sense of group cohesion is critical for successfully implementing new technologies: when individuals care about security, the unwillingness or inability of other group members prevents them from incorporating secure practices in their group~\cite{dencik_data_2016, leistert_resistance_2012}.
Conversely, groups in which individual members lack a commitment to each other or that have unstable memberships struggle to change the ways their members communicate~\cite{agarwal_grassroots_2014}.
{\em We expect that the presence of group motivations to be important for adopting and  continuing to use encryption technologies.}

\paragraph*{Secure Community}
Secure technologies require more than an individual commitment.
{\em We expect that individual users are more likely to adopt a technology when they have a {\bf community} that also uses the technology.}
For encrypted communication, the access to a community is partly due to the network effect~\cite{dingledine_anonymity_2006}. For the network to stay private, all parties must also be using encryption correctly, thus the encrypted communication is only as strong as the weakest user. Research has also found that if the user knew many people were using a secure tool \cite{abu-salma_obstacles_2017, luca_expert_2016}, as well as whether or not there were even a minimal number of influential (e.g. spouse, family members) people using the tool, then this affected their decision to use the tool~\cite{luca_expert_2016}.

Additionally, social influence within a community affects a user's adoption of encrypted technology through those around them using or talking about the technology~\cite{rogers_digital_2017, malhotra_extending_1999}, and in the case of secure technology specifically, through cautionary tales and advice from friends, family, and coworkers~\cite{das_effect_2014}.
For people who lack technical confidence, belonging to a network of encryption-users provides a social resource for troubleshooting. Gaw et al.\ made similar observations in a non-violent, direct-action organization that used email encryption~\cite{gaw_secrecy_2006}.

However, given the timing of our study, we expect that an individual belonging to a community that uses secure technologies will be correlated with not using PGP.
While our PGP trainings started prior to the widespread release of Signal (in late 2015~\cite{greenberg_signal_2015}) and other secure messengers, our survey was deployed in early 2018, well after widespread adoption of secure messenging apps.  {\em We expect that even if  an individual  starts using PGP, they may stop upon their community's adoption of another easier, secure communications technology.}

\paragraph*{Risk}
Many activists are aware of surveillance threats that might derail their planned actions, or increase the severity of the consequences, but the consequences of surveillance vary substantially.
In interviews with activists, Dencik's respondents emphasized that their activism was oriented to the public, so the benefits of being transparent in their planning and communication served a reason to not take up encryption~\cite{dencik_advent_2017}.
Given free-speech protections of most developed democracies (such as the US, where our research is based), activist groups engaged in low-risk activities are likely less worried about surveillance, and thus may be less motivated to protect their communications with each other.
In contrast, higher-risk activism includes activities such as delegitimizing the state, direct-action interference and acts of civil disobedience that are more likely to end in arrest or serious sanctions.
Such activism was subject to state suppressions during the COINTELPRO era~\cite{cunningham_there_2004}, and still is today~\cite{potter_green_2011, marczak_when_2014, marczak_social_2017, marczak_hacking_2015, deibert_communities_2014, scott-railton_security_2016}.  The successful planning and carrying out of high-risk activism requires greater concealment, and it seems likely that individuals engaged in these activities would be more likely to make encryption a continuing part of their organizing strategy, as Gaw et~al.\ observed in their study~\cite{gaw_secrecy_2006}.  That is, at the outset of this work, {\em we expected that level of risk be positively correlated with continued use of encryption.}  However, our analysis uncovers a contraindicative relationship.

\bigskip

In summary, we expect that ease in adopting new technologies or technological confidence, motivation (either individual or group), membership in a community that uses secure technologies, and risk level to have an impact on the adoption and continued use of encryption technologies.
We seek to understand each condition separately, as well as how these conditions might combine to explain long-term use of PGP email encryption.
We do so using qualitative comparative analysis, as described in Section~\ref{sec:qca}.
We describe how we measure the above factors in Section~\ref{sec:measure}.

\section{Methodology}
Prior to the development of our research questions, a subset of our research team had developed and deployed digital security workshops for a local non-profit that provides legal support to social movement groups.
These workshops included both general workshops on surveillance risks and specific, hands-on workshops, including one for PGP email encryption in which
the training team helped workshop participants get set up with PGP email encryption using Thunderbird with the Enigmail plugin (herein referred to as Thunderbird+Enigmail). Thunderbird+Enigmail was chosen by the training team because of its inter-operability, active development and robustness.
The PGP encryption training works through a worksheet (available in the Appendix) with 20 steps, with different worksheets for each of the main operating systems (Mac, Windows, Linux).  The workshops ranged from one-on-one (often over phone or video call) to in-person workshops with as many as 8 participants to 1 trainer.  Across all workshops over 100 participants used the appended worksheet and only one participant failed to send an encrypted email by the end of the workshop.

Our follow-up survey (available in the Appendix) evaluated whether PGP workshop attendees still encrypted their email, along with a variety of other conditions that may shape their decision to adopt encryption following a training session.

\subsection{Recruitment}\label{sec:rec}
Workshop participants were recruited in the following ways: activists and social movement organizations requesting email encryption trainings through the CLDC; paid student workers or volunteers prepared to carry out drop-in center digital security trainings at a regional conference that activist groups attend; and, personal or professional acquaintances of the training team.  None of the participants of the workshops were current users of PGP.  All of the workshops took place before this follow-up research study was conceived.

In early 2018, we sent emails to participants of the PGP workshops for whom we had contact information to request their participation in our research survey (available in the Appendix).  This was 6 to 40 months after participants were trained in Thunderbird+Enigmail.
Of 71 invited individuals, a total of 26 respondents returned at least partial surveys, leaving us with a response rate of 37\% (an acceptable response rate for email recruitment~\cite{keller_what_2014}).
Of these, three were removed from the final analysis because their surveys were missing information used for key conditions or the outcome.
Another four were dropped because they had been paid computer science student workers, recruited to teach others, required to complete the workshop {\em and} received more in-depth training on how to teach others to encrypt emails.  These respondents are qualitatively different from other respondents who voluntarily sought encryption training.

Our final set of respondents includes 19 individuals, which is similar in size to previous PGP evaluations \cite{whitten_why_1999,ruoti_why_2015,sheng_why_2006,garfinkel_johnny_2005}.  We have assigned randomly chosen English words to each individual to refer to them throughout the remainder of the paper.

Nine participants identified as masculine, eight as feminine, and two as gender non-conforming or gender-queer.
One participant identified as both Indian and white, while the rest of the participants were white.
The participants' ages skewed young, with a majority falling between 20 and 39 (see Table~\ref{tab:demo-ages}).
All participants were on the political left, ranging from center left to far left, as self-identified in the survey.  We address the limitations resulting from this biased recruitment in Section~\ref{sec:lim}.

\begin{table}[h]
  \centering
\caption{Participant Ages}
\label{tab:demo-ages}
\begin{tabular}{lllllll}  \toprule
Age & 20-29 & 30-39 & 40-49 & 50-59 & 60-69 & 70+  \\ \midrule
N   & 3     & 11    & 1     & 1     & 1     & 2    \\
\%  & 16\%  & 58\%  & 5\%   & 5\%   & 5\%   & 11\% \\ \bottomrule
\end{tabular}
\end{table}

\subsection{Ethical Principles}\label{sec:eth}
Our research was approved by our Institutional Review Board under exempt category 2. The population we recruit from is one with unique risks, and we aimed to preserve their privacy as much as possible.  The CLDC consults several smaller, independent social movement groups and organizations, and it is these independent groups from which we sampled. These groups are targeted by both law enforcement and groups with opposing ideology, and we take care to maintain trust and to avoid undue harm to these vulnerable groups participating in this research.

We created pseudonymous links to the Qualtrics survey; only the study staff held the mapping to real identities. The initial PGP email workshops avoided collecting attendance information from participants and any contact information was given completely voluntarily. This was done to minimize risk, as well as to assure the community that their confidentiality would be preserved. We borrow these practices from the field of sociology, where certain data collecting limitations are accepted in order to protect, and establish trust with, vulnerable communities.

We offered the recruitment sample a PGP refresher training one month after survey deployment.

\subsection{Qualitative comparative analysis}\label{sec:qca}

In order to understand the pathways leading to (or away from) the adoption of PGP, we employ a Qualitative Comparative Analysis (QCA).  Given the small sample size and the exploratory nature of this work, a qualitative approach best utilizes our rare data, over say, a quantitative statistical analysis.  This choice of analysis also reflects the interdisciplinary nature of our collaboration (between sociology and computer science) that aims to ultimately tease apart the group dynamics involved in adopting secure communications technology.

QCA was developed as a method to infer the configuration of causal\footnote{The literature around QCA anecdotally uses the term {\em causal}, while the true relationship is generally correlational.} factors that lead to a given outcome and is generally used to analyze a smaller number of cases for which many details are known (for example, nation-states). While QCA is {\em not} a quantitative method, it brings standardized rigidity and reproducibility to qualitative analysis that helps organize patterns and correlations among themes. While our cases are a far cry from those of nation states, more recent work has successfully applied the method to individual respondents~\cite{ragin_intersectional_2016,warren_applying_2013,borgna_multiple_2016} and QCA has seen widespread use beyond political science~\cite{roig-tierno_overview_2017}.  Refer to Ragin's book {\em Redesigning Social Inquiry} for a thorough background of QCA methodology~\cite{ragin_redesigning_2008}.  QCA best practices suggest that QCA is appropriate for our number of cases (19) and level of detail about each case~\cite{rihoux_configurational_2009}.  QCA has been applied to datasets of similar size and level of detail, including studies of social movements~\cite{mcadam_site_2010, giugni_paths_2013}.

In the {\em crisp set} form of QCA, each case is described by a set of binary conditions and a binary outcome.  (Our cases and their binary descriptions are given in Table~\ref{tab:truth} with each condition denoted by an all-caps name.)  Crisp-set QCA uses Boolean minimization (via the Quine-McCluskey algorithm or equivalent~\cite{dusa_qca_2019}) to find a Boolean formula in disjunctive normal form of minimum size that infers the outcomes.  The Boolean minimization acts over the complete truth table of condition combinations with each row having a corresponding (binary) outcome.   Generally, different formulae are uncovered to infer the positive outcomes separately from the negative outcomes (as we will discuss further in Section~\ref{sec:asym}).  Each clause in a formula is referred to as a {\em causal pathway} in the QCA literature. While there are other forms of QCA, such as ``Fuzzy-Set'' which uses sliding scale values to organize outcomes, we employ the ``Crisp-Set'' (binary) method, which is more suitable to the size of our data set and the nature of our conditions, most of which were naturally binary or easily dichotomized, as we describe in Section~\ref{sec:measure}.

However, the given cases may not cover all possible combinations of conditions, and may indeed only represent a fraction of the rows of the complete truth table (referred to as {\em limited diversity}).  Further, multiple cases may have the same combination of conditions but have conflicting outcomes.  QCA handles the configurations for which we have no cases (known as {\em remainders}) of conflicting cases in three basic ways, resulting in three types of QCA solutions~\cite{thiem_boolean_2013}.  We describe these in terms of inferring the positive outcome; symmetric processes and definitions hold for the negative outcome~\cite{rihoux_configurational_2009}.

In the {\em conservative solution}, Boolean minimization is
  performed over the truth table composed of all the positive
  outcomes, and treats all the remainders as well as negative and conflicting
  outcomes as negative.  The resulting Boolean formula evaluates to
  true {\em only} for the combinations of conditions for which all the cases
  have positive outcomes.

 In the {\em parsimonious solution}, Boolean minimization is performed over the truth table composed of all the positive outcomes, treats all the negative and conflicting outcomes as negative, and treats all the remainders as {\em don't cares}.  The resulting Boolean formula evaluates to true for the combinations of conditions for which all the cases have positive outcomes, evaluates to false for the combinations of conditions for negative or conflicting cases, and evaluates to true or false for the remainders.

 An {\em intermediate solution} lies between the conservative and parsimonious solutions and is obtained by making hypotheses about the direction in which literals should influence the outcome~\cite{ragin_redesigning_2008}.

For all three solutions, a conflicting outcome may be treated as either positive or negative via a threshold, for example as positive if at least 70\% of the cases are positive.  Further, one may relax how perfectly any given pathway infers positive outcomes, allowing a pathway to capture some small fraction of negative outcomes, a notion formalized by consistency.

\begin{table*}[t]
  \centering
  \caption{Truth Table: Cases and their conditions.  Note that Confident and Kindred have the same set of conditions but have a conflicting outcome (in this case, ENCR$_{\mbox{Confident}}$ = 1 and ENCR$_{\mbox{Kindred}}$ = 0).  Condition definitions are given in Section~\ref{sec:measure}.}
  \label{tab:truth}
  \setlength\tabcolsep{1.5pt}
  \begin{tabular}{c|c|c|c|c|ccl}\toprule
    \multicolumn{5}{c}{Conditions} & Outcome & \multicolumn{2}{c}{Cases} \\\cmidrule(lr){1-5}\cmidrule(lr){6-6}\cmidrule(lr){7-8}
    EASE	&	MOT-I	&	MOT-G	&	COMMS	&	RISK	&	ENCR	&	\#	& pseudonyms	\\ \midrule
0	&	0	&	1	&	1	&	1	&	0	&	3	&	Cranial, Sitter, Passover	\\
0	&	1	&	0	&	0	&	0	&	1, 0	&	2	&	Confident, Kindred	\\
0	&	1	&	0	&	0	&	1	&	0	&	1	&	Afterlife	\\
0	&	1	&	0	&	1	&	1	&	0	&	1	&	Gracious	\\
0	&	1	&	1	&	1	&	0	&	0	&	1	&	Resent	\\
1	&	0	&	1	&	0	&	0	&	1	&	1	&	Olive	\\
1	&	0	&	1	&	0	&	1	&	0	&	1	&	Evacuee	\\
1	&	0	&	1	&	1	&	0	&	1	&	1	&	Resale	\\
1	&	0	&	1	&	1	&	1	&	0	&	1	&	Handmade	\\
1	&	1	&	0	&	1	&	0	&	1	&	1	&	Democracy	\\
1	&	1	&	0	&	1	&	1	&	1	&	3	&	Cargo, Drab, Problem 	\\
1	&	1	&	1	&	1	&	1	&	1	&	3	&	Sixtieth, Snowsuit, Vendetta 	\\\bottomrule
  \end{tabular}
\end{table*}

\paragraph*{Consistency}\label{sec:con} For a logical formula, {\em consistency} is the fraction of cases satisfying the logical formula that have positive outcomes.  For example, of our 11 cases with the condition EASE=1 (see Table~\ref{tab:truth}), 9 continue to use PGP (ENCR=1).  Therefore, $\mbox{EASE} \Rightarrow \mbox{ENCR}$ has consistency $9/11 = 0.82$.  Consistency scores of individual conditions (or their negations) for our data set are given in Table~\ref{tab:con-cov}.  QCA scholars use consistency scores of $\ge 0.8$ as goals for analysis~\cite{ragin_redesigning_2008}.  Indeed the pathways we uncover all achieve perfect consistency ($=1.0$), and so we are able to avoid relaxing the constraints for Boolean minimization described above.

\paragraph*{Coverage}\label{sec:cov}{\em Coverage} is of secondary importance to consistency.   The coverage of a logical formula is the fraction of cases with positive outcome whose conditions satisfy the formula.  For example (see Table~\ref{tab:truth}), of the 10 cases who continue to use PGP (ENCR=1), 9 have the condition EASE=1.  Therefore, $\mbox{EASE} \Rightarrow \mbox{ENCR}$ has coverage $9/10 = 0.9$.  Coverage scores of individual conditions (or their negations) are given in Table~\ref{tab:con-cov}.  There are no  thresholds for coverage established in the literature, but generally, the higher the value, the more relevant a logical formula is for explaining the occurrence of an outcome.

\bigskip

\noindent In this work, we use the {\em parsimonious solution}.  In analyzing the parsimonious solution, we will discuss the remainders (configurations for which we have no cases) that are covered by the pathways of the resulting parsimonious solutions (Section~\ref{sec:remain}).  Given our limited cases, the pathways of even the intermediate solutions are either very similar to those of the parsimonious solution or have such low coverage as to only explain one or two cases, making the parsimonious solution more useful at this stage of research.

We use the software fsQCA 3.0 for the QCA-style Boolean minimization~\cite{ragin_fuzzyset_2016}.

\subsection{Measurement of the conditions} \label{sec:measure}
We measure the conditions we hypothesize to be important for adopting PGP email encryption as binary variables reflecting the presence or absence of that condition.  Our conditions are either dichotomous by nature or easily dichotomized via thresholding at a naturally occurring gap in a scaled response, according to the best practices of QCA methodology.  The conditions were derived from responses on the survey and the in-depth knowledge the training team had of the survey participants.  We define the conditions below and explain how we derive the measured conditions from the original responses on the survey and give examples of relevant qualitative responses.  The measurements for our 19 cases are given in Table~\ref{tab:truth}.  The pseudonym for each case is a randomly selected word.

\paragraph*{Continues to use PGP email encryption---ENCR}
The outcome condition was a simple binary value indicating whether the respondent was still using a PGP email encryption system at the time of the follow-up survey.  The outcome condition was coded as true if the respondent replied ``yes'' to at least one of the following two survey questions:
\begin{enumerate}
    \item Do you continue to use Thunderbird+Enigmail?
    \item Do you use an email encryption system other than Thunderbird+Enigmail?
\end{enumerate}
For the two respondents who selected ``no'' for the first question and ``yes'' to the second, we confirmed that the email encryption system they referred to in question 2 was indeed a PGP email encryption system through their answer to the survey question ``Which system, and why do you use it? Please list and discuss all systems you use and why.''  (In both cases, the respondents indicated they adopted ``Mailvelope'' sometime after the Thunderbird+Enigmail training.)

\paragraph*{Technological Ease---EASE}
In our survey, we included a subset of the validated System Usability Score (SUS) questions~\cite{brooke_sus_1996} for Thunderbird+Enigmail and computer self-efficacy questions~\cite{compeau_computer_1995}.  We found little variation in the self-efficacy responses among our study participants, giving little discriminatory power for  QCA.  However, among the usability questions, we found sufficient variation to provide a useful condition for QCA.  We used the responses on
six questions about study participants' experience being trained on and using Thunderbird+Enigmail:
\begin{itemize}
\item I found Thunderbird+Enigmail unnecessarily complex.
\item I thought Thunderbird+Enigmail was easy to use.
\item I think that I would need the support of a technical person to be able to use Thunderbird+Enigmail.
\item I found Thunderbird+Enigmail cumbersome to use.
\item I felt confident using Thunderbird+Enigmail.
\item I need to learn a lot of things before I could get going with Thunderbird+Enigmail.
\end{itemize}
Respondents were asked to indicate their level of agreement with these statements on a five point Likert scale with 1 being strongly disagree and 5 being strongly agree.
All questions were coded or recoded so that a score of 5 indicated more technical confidence or ease of use with technology.
Scores were then combined to give each respondent a raw score between 6 and 30.
Eight cases fell at 18 or below; 11 fell at 20 or above.
The score 19 offers a natural cut-off point between the lower group and the higher group:
we code a participant with EASE=1 if they have a score of 20 or above.

\paragraph*{Motivation for Learning Encryption---MOT-I \& MOT-G}
Respondents were asked the open ended question, ``Why did you want to learn to use email encryption? What was your reason for the training?''
We used responses to these to code whether respondents were motivated to participate in a PGP email workshop for personal (individual) or communal (group) reasons.
Responses were inductively coded along themes that emerged in their responses~\cite{strauss_grounded_1997}, using true to represent the presence of the given type of motivation:
\begin{itemize}
\item{\em MOT-I:} Respondents with {\em individual motivation} present were those whose responses included references to their personal concern about surveillance from the state or from corporations or curiosity.  For example: ``I was curious about it and the instructors offered a free lesson while I was at work.  Thought it might be useful in the future if I needed to feel secure that I was communicating confidentially.'' and ``I wanted to understand the technology and know how to use it so that I would be able to.''
\item{\em MOT-G:} Respondents with {\em group motivation} present were those whose responses included references to planning direct-action activism, a desire to keep group communication secure, or the desire to provide cover to activists.  For example: ``To be able to institute the practice in my organizing collective.'' and ``In case some direct action needed to use more secure email.''
\end{itemize}
These categories were not mutually exclusive; respondents could be coded as having both conditions simultaneously.
We expect that the presence of motivation (either group or individual) will be important for continued use of PGP email encryption.

\paragraph*{Community Using Secure Technology---COMMS}  Respondents were asked the yes/no question ``Are you part of a community of people who use technologies for secure communication?'' We coded the responses accordingly.

\paragraph*{RISK}
It would pose needless hazard to the human subjects of our research to collect as part of the survey their self-reported risks they face through their activism, and it may dissuade participants from completing their response.
In order to evaluate the risk of the subjects' social movement activities,
the training team member who conducted all of the PGP trainings assessed each participant for the risk they faced of arrest and/or other sanction.
These assessments were based on the trainers' deep knowledge of each respondent's activism.
We coded individuals as high risk (RISK=1) if their activities could result in measurable negative consequences given current US laws and practices (from loss of employment to fines or criminal charges).
Remaining individuals were coded as low (or no) risk (RISK=0).
While we hypothesized that risk be positively correlated with continued use of encryption technologies, our QCA has borne out a contraindicative relationship, and so we include both positive and negative correlations in Table~\ref{tab:con-cov}.

\bigskip

\noindent We describe above only those conditions for which we have hypotheses that were borne out in the data.  We investigated other conditions (such as nature of and time since training, and demographics) using QCA and found no discernible pattern.  We describe this further in Section~\ref{sec:lim}.  In our survey, we asked ``Are you part of a community of people who use email
encryption?'' We were surprised that this condition did not present in pathways through QCA.  It may be that respondents interpreted this question in different ways (temporally, for example).  Or it may be that network and community affects highlighted by QCA were more usefully encoded by RISK or COMMS. However, the question did fill the role of {\em attention check} with everyone answering yes to this question also answering yes to ``Are you part of a community of people who use technologies for secure communication?''

\section{Results}
\begin{table}[h]
  \centering
  \caption{Consistency and coverage scores of conditions for which we have defensible hypotheses}
  \label{tab:con-cov}
  \begin{tabular}{rrrcrrr}\toprule
    & \multicolumn{2}{c}{$\Rightarrow$ENCR} && & \multicolumn{2}{c}{$\Rightarrow\neg$ENCR}\\\cmidrule(lr){2-3}\cmidrule(lr){6-7}
    & cons. & cov. &&& cons. & cov. \\\cmidrule(lr){1-3}\cmidrule(lr){5-7}
    EASE  & 0.82 & 0.90 && $\neg$EASE & 0.88 & 0.78 \\
    MOT-I & 0.67 & 0.80 && $\neg$MOT-I & 0.71 & 0.56  \\
    MOT-G & 0.45 & 0.50 && $\neg$MOT-G & 0.38 & 0.33 \\
    RISK  & 0.46 & 0.60 && $\neg$RISK & 0.33 & 0.22 \\
    $\neg$RISK & 0.67 & 0.40 && RISK & 0.54 & 0.78 \\
    & &                       && COMMS & 0.43 & 0.67 \\ \bottomrule
  \end{tabular}
\end{table}

\subsection{Coverage and consistency of conditions}\label{sec:con-cov}

In Table~\ref{tab:con-cov}, we give the consistency and coverage of each hypothesized causal condition for each outcome of interest (ENCR=1 and ENCR=0), as calculated using the formulae outlined in Section~\ref{sec:qca}.
Note that we do not include COMMS or $\neg$COMMS as a condition for the positive outcome, as we do not have a hypothesis for this in the causal pathway for the positive outcome as earlier discussed.
We include all directions of correlation for RISK, as the result has been counter to our initial hypothesis regarding RISK.

Only EASE and $\neg$EASE reach the 0.80 threshold for consistency for ENCR and $\neg$ENCR, respectively.  In this light, we should expect that technical confidence is likely to be relevant within the causal pathways in the next set of analyses.
Whatever other conditions play a role in continued encryption use, they are likely to build on existing technical confidence and ease in respondents.
This makes sense given the technical requirements of PGP email encryption that require tasks not usually associated with email or other common programs.

Note that RISK is more consistent with $\neg$ENCR than $\neg$RISK, and, likewise, $\neg$RISK is more consistent with ENCR than RISK,  giving the first indication that the correlation of RISK with ENCR is negative and seemingly
contradictory, or at least certainly opposite to what security professionals would hope---that individuals at higher risk would take more precautionary steps.

\subsection{Truth table and diversity of cases}\label{sec:table-obs}

In Table~\ref{tab:truth}, we present the binary conditions of our 19 cases, collapsed to common combinations of conditions.  Only two cases with a common combination of conditions (participants Confident and Kindred) conflict in their outcome, with Confident continuing to use PGP and Kindred not.

The outcomes are quite balanced, with 10 individuals continuing to use PGP and 9 not.
This makes our case set very suitable for QCA analysis.
This is also a quite different outcome from PGP usability studies in which participants are rarely able to successfully encrypt a single email.

We note that while our cases only cover 12 of the 32 possible combinations of conditions, all but one of the remainders fall into (partially overlapping) groups with accompanying explanations:
\begin{itemize}
\item All 8 condition combinations corresponding to MOT-I=0 and MOT-G=0 are remainders.  An individual with neither individual nor group motivation is very unlikely to participate in a training.  We expect such individuals to be unlikely to adopt a complex tool such as PGP, but more accurately this combination (low motivation) of conditions is simply out of the scope of the present research.
\item 6 of the 8 condition combinations corresponding to RISK=1, COMMS=0 are remainders.  All the high-risk participants in this study would be aware of state and corporate surveillance as part of our training. With the prevalence of secure messengers by the time of the survey and our experience with supporting the digital security needs of direct-action activists, we believe that most high-risk activists adopt some sort of secure communications.  Indeed, the two cases with RISK=1, COMMS=0 (participants Afterlife and Evacuee) are engaged in social change through their employment, whereas our other participants with RISK=1 are engaged in grassroots social change.  Afterlife and Evacuee may not have control over the adoption of E2EE technologies in their employment.
\item 11 of the 16 combinations with RISK=0 are remainders.  Many of our PGP workshop attendees seek out training through or are recommended for training by our collaborating non-profit that provides legal support for front-line activists.  This non-profit is more likely to recommend digital security training such as our PGP workshop to activists who are in groups who have required legal support, indicating higher risk, in order to protect legal communications.
\end{itemize}
The remainder that doesn't fall into these groups is: EASE=0, MOT-I=1, MOT-G=1, COMMS=1, RISK=1.

\subsection{QCA for the positive outcome}

The parsimonious solution for the positive outcome (ENCR=1) uncovers the following logical formula:
\begin{equation}
  \label{eq:pars-encr}
  (\mbox{EASE}\land\mbox{MOT-I}) \lor (\mbox{EASE}\land\neg\mbox{RISK}) \Rightarrow \mbox{ENCR}
\end{equation}
This logical formula has perfect consistency (all the cases which satisfy the antecedent have ENCR=1) and coverage 0.9 (of the 10 cases with ENCR=1, 9 satisfy the antecedent of Formula~\eqref{eq:pars-encr}).
The logical formula only fails to cover the case of participant Confident who continues to use PGP; however, Confident conflicts with participant Kindred---therefore, it is impossible to uncover a formula which achieves both perfect consistency and perfect coverage.

Formula~\eqref{eq:pars-encr} is equivalent to:
\begin{equation}
  \label{eq:pars-encr-2}
  \mbox{EASE}\land (\mbox{MOT-I} \lor \neg\mbox{RISK}) \Rightarrow \mbox{ENCR}
\end{equation}
which highlights that while technical ease (EASE) is necessary for both causal pathways, it is not sufficient, and that it needs to be combined with either individual motivation (MOT-I) or low risk ($\neg$RISK).

The first causal pathway, $\mbox{EASE}\land\mbox{MOT-I}$, requiring that technical ease be combined with personal motivation, is consistent with both the literature and our expectations that most people who take up higher-level security measures need to have at least some technical confidence, and need to be fairly concerned about the risks of surveillance for their personal information.
It is also likely that many of those with higher-level technical skills and confidence are more aware of surveillance risk, even when compared with other respondents who also sought out encryption training.
This causal pathway achieves coverage 0.7.

The second causal pathway ($\mbox{EASE}\land\neg\mbox{RISK}$) covers the cases of participants Olive, Resale, and Democracy (coverage 0.3), with Democracy also satisfying the first causal pathway.  In an intermediate solution, this pathway also includes MOT-G; in fact, $\mbox{MOT-G}\land\neg\mbox{RISK}$ only implies ENCR when the individuals also had technical ease.  This occurs in two cases (Olive and Resale), both emerging from a single informal group of retired friends who sought training together.
Of the three members of the low-risk group that sought training, only two continued to use PGP.
A third member (Resent) also reported MOT-G but not EASE, and this member was the only one to discontinue using PGP.
This underscores the importance of technical comfort as an issue that must be dealt with among those advocates seeking to increase the use of encryption.

This second pathway provides useful insight for the surprising role of risk in predicting individual behavior.  We revisit this in Section~\ref{sec:risk}.

\subsection{QCA for the negative outcome}

QCA does not assume symmetry when predicting an outcome versus predicting its absence: we separately assess combinations of conditions leading to discontinued use of PGP encryption.
This also provides a check on the relevance of the conditions in the causal pathways to the continued use of encryption.
As we noted above, we can be more confident that the causal paths we have discovered are meaningful if we can also demonstrate that their absence is important when explaining the outcome's absence.

The parsimonious solution for the negative outcome (ENCR=0) uncovers the following logical formula:

\begin{equation}
  \label{eq:pars-nencr}
  \begin{split}
  (\neg\mbox{EASE}\land\mbox{COMMS}) \lor (\neg\mbox{EASE}\land\mbox{RISK})&\\\lor (\neg\mbox{MOT-I}\land\mbox{RISK}) &\Rightarrow \neg\mbox{ENCR}
\end{split}
\end{equation}
This logical formula has perfect consistency (all the cases which satisfy the antecedent have ENCR=0) and coverage 0.89 (of the 9 cases with ENCR=0, 8 satisfy the antecedent of Formula~\eqref{eq:pars-encr}).  The logical formula only fails to cover the case of participant Kindred who discontinues PGP use.  However, Kindred conflicts with participant Confident. Therefore, as with the pathway for the positive outcome, it is impossible to uncover a formula which achieves both perfect consistency and perfect coverage.

We can immediately observe that the factors that lead to the adoption of PGP play a role in the discontinued use of PGP in their negated form, confirming the relevance of the conditions EASE, RISK and MOT-I.

Each clause in formula~(\ref{eq:pars-nencr}) achieve a coverage of 0.56 with 5 of the 9 cases satisfying each clause.  As a result these pathways have a high degree of overlap.
Participants Passover, Cranial and Sitter satisfy all three pathways; participant Gracious satisfies the first two pathways.

Formula~\eqref{eq:pars-nencr} highlights the importance that a lack of technical ease (EASE=0) and a higher-risk persona (RISK=1) have to discontinued use of PGP email encryption.  Given that EASE is pivotal to continued use of PGP (ENCR=1), it is natural that a lack of technical ease is important to discontinued use of or failure to adopt PGP.  That RISK helps to explain 6 cases (pathways 2 and 3), we have additional proof that it is risk aversion that drives a continued commitment to encryption, rather than a sense that respondents have an objective risk level that requires greater security.

In the third pathway for the negated outcome, a lack of individual motivation is combined with  risk ($\neg\mbox{MOT-I}\land\mbox{RISK}$).
This makes some theoretical sense, given that those who are the least worried about security in their communications are likely to be those who are willing to take some larger risks in their activism.

The first pathway for failing to adopt PGP email encryption uncovers the role played by having a community who use secure communications technology.  Notably, if an individual both belongs to a community using a E2EE communications tool such as Signal {\em and} they do not have technological ease, they fail to adopt PGP email encryption.  This confirms our hypothesis regarding COMMS along with a natural combination with $\neg$EASE.

We asked our respondents: ``If you no longer use Thunderbird+Enigmail, why?''  Two individuals cited technical problems (both with EASE=0):  participant Cranial switched to a new computer and failed to set PGP up again and participant Passover forgot ``how to do the key thing with other users.'' Participant Evacuee wanted cross-compatibility with their smartphone.  Six individuals referred to leaving the GMail interface as being the main barrier (including the two users who switched to Mailvelope to access PGP): ``Hooked on the convenience of non-encrypted Gmail.'' (participant Resent).  Three individuals cited a lack of need.  Participant Sitter noted ``I have found something that is easy to replace email [with] is signal.''  Only one respondent (Handmade) echoed the fatalist opinion found widespread among leftist groups post Snowden~\cite{dencik_data_2016}, saying ``The vast majority of my emails are not activist related and my online, financial, and public behavior, if a law enforcement agency wanted them, would be public.''

\subsection{Asymmetric pathways}\label{sec:asym}

As you can see from Formulae~(\ref{eq:pars-encr}) and~(\ref{eq:pars-nencr}), QCA returns asymmetric pathways to explain positive and negative outcomes.
This is natural as positive outcomes generally have simpler explanations than negative outcomes: there are often many more ways and reasons to stop doing something than to continue.  The pairs of positive and negative pathways, interesting in their own right, also lend extra credibility to those conditions that appear in both pathways.  Indeed those conditions appearing in both pathways (EASE, RISK, MOT-I) occur negated in the opposite pathway.

\subsection{Covered remainders} \label{sec:remain}

\begin{figure}
  \includegraphics[width=\textwidth]{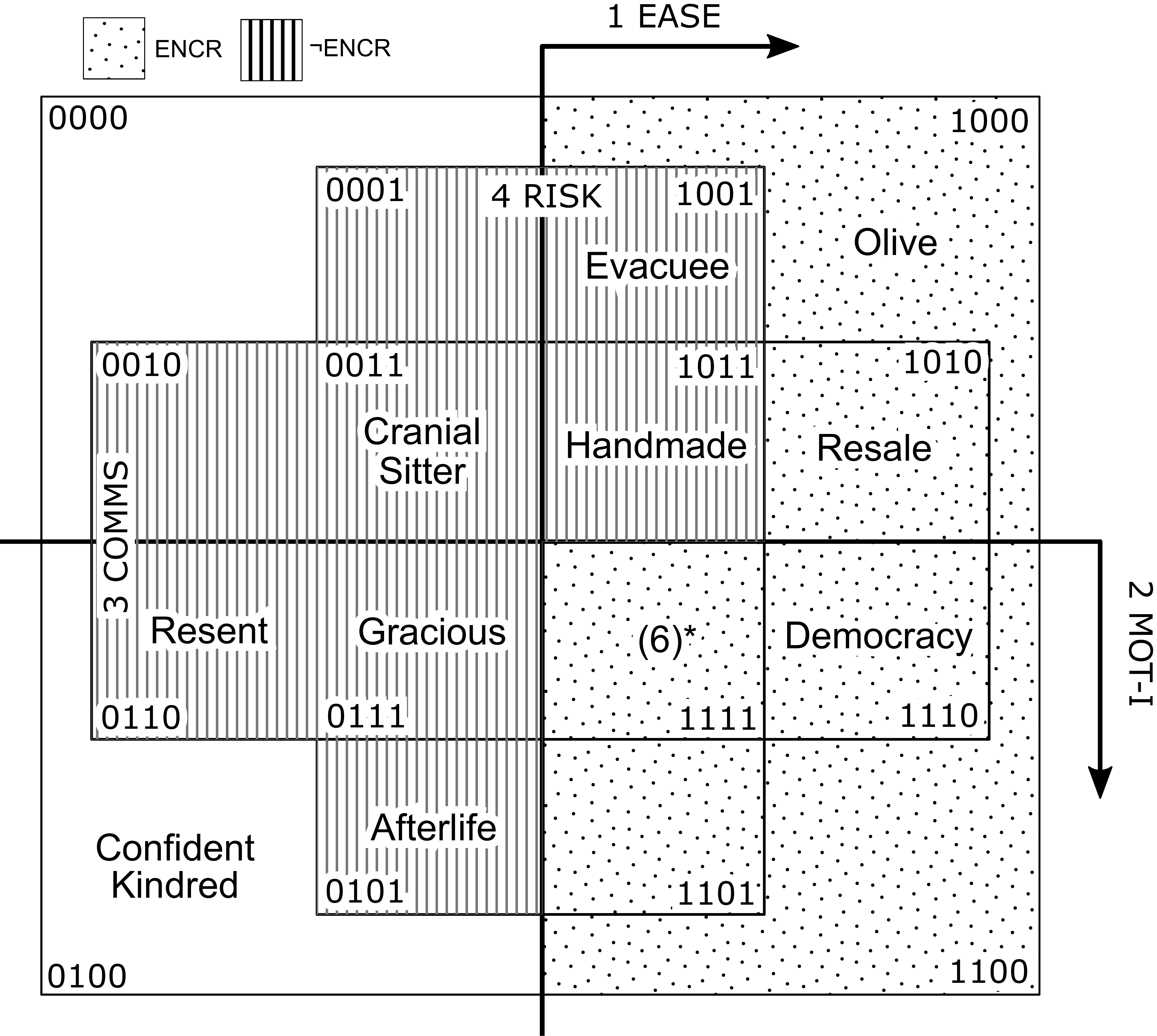}
  \caption{A Venn diagram of the condition space over the conditions: EASE, MOT-I, COMMS and RISK.  The combinations of conditions captured by the parsimonious solution for ENCR (Formula~(\ref{eq:pars-encr})) is given by the dotted area while the striped areas are the combinations of conditions captured by the parsimonious solution for $\neg$ENCR (Formula~(\ref{eq:pars-nencr})). (6)* refers to the six participants: Cargo, Drab, Problem, Sixtieth, Snow, and Vendetta.}
  \label{fig:venn}
  \vspace{-2em}
\end{figure}

In Figure~\ref{fig:venn}, we illustrate Formulae~(\ref{eq:pars-encr}) and~(\ref{eq:pars-nencr}) in the Venn diagram of the case space.  Since MOT-G did not appear in the pathways of the parsimonious solutions, we omit MOT-G from this illustration.  This highlights those combinations of conditions for which our parsimonious solutions are making predictions, for example EASE=1, MOT-I=1, COMMS=0, RISK=0 (in the bottom right).  These are potentially interesting cases that we haven't observed, and indeed, as discussed in Section~\ref{sec:table-obs}, may not be possible to sample.

\section{Discussion \& Related Work}\label{sec:discuss}

The literature on PGP email encryption use has consistently found that it has poor learnability in laboratory settings.
We test these findings with individuals in their natural social context, assessing how this context might make encryption use more likely.
Indeed, we find that the rate of long-term PGP use by our respondents is over 50\%, seeming to counter the poor learnability uncovered by laboratory studies.  Indeed, in our original PGP email workshops, all but one participant were able to successfully send and receive PGP-encrypted emails.

Renaud, Volkamer, and Renkema-Padmos identify a sequence of seven steps that users must overcome in order to uptake E2EE from understanding through ability and deployment~\cite{renaud_why_2014}.  Our respondents who did not engage in long-term PGP use (and surveyed months after the workshop) are still making it through to step five (``Knows E2EE Helps'') or six (``Able to Use E2EE'') based on the survey responses, but are not making it to the seventh and final step ``Is Not Side-Tracked'': They are able to successfully use PGP email at least once, but still fail to adopt in the long-term for another reason.

While our research is unique in trying to understand the social factors that lead to success in that last step, there are two earlier papers with which our work dovetails.

Gaw, Felton, and Fernandez-Kelly performed an ethnographic study of a non-violent, direct-action social movement organization (``ActivistCorp'') that used PGP email~\cite{gaw_secrecy_2006}.  While they find that members of ActivistCorp manage to adopt PGP, it is not without complaint.  Counter to the respondents of our survey, ActivistCorp uses internal authority in order to enforce and support the use of PGP by the individuals of the organization.  Despite this, there was variance in the degree to which ActivistCorp members used PGP for internal emails, from encrypting all emails to only encrypting sensitive emails.  Gaw et al.\ found that this variance was explained by both technical and social factors, from annoyance (as there was technical overhead to managing the encrypted emails) to views on secrecy (such as over-encryption indicating paranoia).  There are similarities between the factors that Gaw et al.\ describe and our factors of EASE and MOT-I.
Although we only determine whether or not our respondents are continuing to use PGP (and not the degree to which they are using PGP), we can draw a comparison between ActivistCorp members who use PGP more (and whose comments indicate a higher level of individual motivation and ease of use with the technology) and our respondents who continue to use PGP (and have EASE = 1 and/or MOT-I = 1).

McGregor, Watkins, Al-Ameen, Caine, and Roesner investigated the success of security measures in the Panama Papers project~\cite{mcgregor_when_2017}. Despite collaboration of 354 journalists working remotely across the globe on the same project, the project avoided breaches using tools such as two-factor authentication and PGP email. This is largely accredited to the resources of the organization and the strong security culture instilled on the collaborators.  The project-specific tools are also to credit as they employed user-centered and security-by-design, providing features valued by the collaborators~\cite{norman_user_1986}.  While the resources of the Panama Papers collaboration far exceeds those of the groups our respondents represent, we may be able to infer that the instillation of security culture gave collaborators individual motivation (MOT-I = 1) and that the tailoring of the tools to the users increased the ease of their use (EASE = 1).

\subsection{Group vs.\ individual factors}

Gaw et al.\ and McGregor et al.'s work each focus on a single organization, and so cannot comment on group factors that may come into play, as we begin to with this research since our respondents hail from different organizing groups.  While MOT-G did not play a role in the causal pathways, other group factors did (RISK and COMMS).  RISK is determined by potential consequences of a group's actions and COMMS measures a group's commitment to digital security if they use secure communications for organizing work.  COMMS and RISK may more reliably measure group-based motivation than our coding of responses to the open-ended questions ``Why did you want to learn to use email encryption? What was your reason for the training?'' that was used to encode MOT-G.

However, individual level factors seem to matter most for people when choosing to continue with encryption.
While the role of EASE (which we measure using a subset of the standard usability questions) has been studied before, the interplay of this condition with others has not.
Level of confidence with the technology (EASE) carried the most weight.
Beyond ease with the technology, there were two pathways that lead to continued encryption use, one based only on individual level factors, and one which includes group factors.
At the individual level, respondents needed to feel personally concerned about surveillance.
The second pathway encompassed respondents who belonged to low-risk movement groups, yet their commitments to each other seemed to offer a boost to their individual commitments to using encryption.
These group commitments did not matter for all members equally, however.
Lacking technical confidence seems to cancel out the commitment to group protection.

Existing research indicates that if the functionality of the secure software inhibits or slows down the users' ability to perform essential tasks, users will either elect not to use it \cite{mcgregor_investigating_2015, abu-salma_obstacles_2017, marczak_social_2017}, or if users are required to use secure software (such as through an organizational requirement) users will attempt to work around it \cite{mcgregor_individual_2016, marczak_social_2017}. To further support this, in the project reviewing the success of the release of the Panama Papers, McGregor et al.\ note how users were able to accept security requirements when their general technology needs were addressed by the secure system they were using \cite{mcgregor_when_2017}.

\subsection{The unfortunate role of risk}\label{sec:risk}

It has previously been established that many individuals partake in a privacy paradox: they often have stronger beliefs about privacy behaviors and ideals than their actions would indicate~\cite{norberg_privacy_2007}. However, the phenomenon may not be so paradoxical. A users' privacy concerns may be at odds with their access to, knowledge of, or ability to perform counter measures \cite{kokolakis_privacy_2017}. Renaud et al. establish a sliding scale of interpretations for the privacy paradox that consist of different combinations of level of understanding, access to technology, and ability to perform protective actions \cite{renaud_why_2014}. Abu-Salma et al. highlight the more nuanced obstacles to adoption of secure communication tools, namely that users have knowledge gaps when making decisions about how to protect their communication, and make flawed threat models based on these flawed mental models~\cite{abu-salma_obstacles_2017}.

In this work we find a risk paradox---that risk does not play the role that most security professionals would likely hope: that individuals undertaking riskier activities would be more likely to adopt a higher security profile by adopting PGP email encryption.  Rather, we find the latter, with risk negatively impacting adoption of PGP.
This may be because higher-risk activists were already accustomed to planning riskier actions only in face-to-face meetings, and thus felt no need to incorporate secure communication.
It is also possible that in contrast to the risk-averse respondents, respondents who engage in riskier activism may simply have a higher tolerance for risk in their communication.
One respondent reported their sense that even ``militant'' civil disobedience did not require secure email, and that much of their personal information would be available online to state authorities anyway.
From this perspective, activists may consider increased email security an unnecessary step.  Regardless, the relationship of risk-taking to security practices would be interesting to investigate through a more generalizable study.

\subsection{Limitations} \label{sec:lim}

Our study is retrospective.  The workshops in which participants learned to use PGP email encryption had taken place before this study was designed.  A number of the limitations stem from the difficulty in studying use of secure communications technology in the wild, particularly in a particular population (social movement participants) using a relatively rare technology (PGP email).  This study, though, was made possible through our partnering with the CLDC.  This results in a number of limitations:

\paragraph*{Workshop variation} As we noted, there is a fair amount of variation in the workshops.  However, a number of factors were controlled: the same trainer (a member of the research team) hosted every workshop, Thunderbird+Enigmail was consistently used as the platform, and all trainings followed the same worksheet (see Appendix). Despite this, there was still unavoidable variations: trainer-to-participant ratios ranged from 1:1 to 1:8, workshops could occur as a drop-in center or a fixed-time workshops, and sometimes were conducted over phone or video.

\paragraph*{Time since workshop} We investigated whether time was a factor in measuring continued use of PGP email encryption.  There are two hypotheses for how this time interval could be a confounding factor.  First, the longer the time between a workshop and follow-up, the more opportunities there are to stop using PGP email encryption.  In contrast, the more time passes, the less likely someone may be to respond to a recruitment email, and this may be further confounded by the response biases we describe next.  We investigated time as a factor that could affect our respondents' outcomes and found that time since training did not affect the outcome.

\paragraph*{Response bias}  While all studies that recruit human subjects are vulnerable to response bias, we reflect here on those aspects particular to our study.
A workshop participant who continues to use PGP email encryption may be more likely to respond to our survey as they may have more interest in the study.  Conversely, a workshop participant who does not continue to use PGP email encryption may be embarrassed by this fact and be less likely to respond to the survey.
Participants from two workshops were inaccessible to us due to the way in which the workshop was convened, without the trainer or partnering non-profit having a list of workshop attendees.
These workshop participants were from tightly-knit, highly-private groups that engaged in higher-risk activism.
In these cases, we did attempt to recruit through the workshop convener, but did not receive any responses this way.  As such, we may have lower response rates from higher -risk workshop participants.  This bias is in the opposite direction of that among recruitment for training (as we comment on in Section~\ref{sec:table-obs}), where higher-risk activists are more likely to be recommended for digital security training.

\paragraph*{Left bias} The participants all identify with being on the political left.  This bias is a result of  partnering with the CLDC which supports mostly environmental and social justice groups, which tend to engage left-leaning individuals.  However, the work of Agarwal et al.~\cite{agarwal_grassroots_2014}, which studied the values and technology choices of both left- and right-wing social movements found that security and privacy concerns were similar on both ends of the spectrum.  This may mitigate the concerns of the political bias of our participants.

\paragraph*{Demographic bias} We acknowledge that our participants' demographics are skewed young and white; this reflects the demographics of the population from which we recruit in Oregon, which is approximately 85\% white.

\section{Conclusion \& Recommendations}

Work exploring how motivated users in communities impacted most by surveillance interact with new or novel secure technologies and evaluating usability is useful, such as in the work of Lerner, Zeng, and Roesner~\cite{lerner_confidante_2017}.  Lerner et al.\ sampled lawyers and journalists in order to evaluate the development of Confidante, a prototype PGP email client using Keybase.io's key management to address key exchange issues. However, Confidante is still restricted to use in a lab setting, and the project is still in beta with no planned release as of this writing, restricting its accessibility for users with immediate needs.  Gaw et al.'s study of ActivistCorp looked at how an at-risk organization adopted existing (and difficult-to-use) technology.  While
their contributions cannot be understated, significant exposures about surveillance have come to light since their 2006 paper, as well as significant technological progress, which may limit its continued relevance. Our work follows in the footsteps of Gaw et al., seeking to
understand how at-risk communities interact with and adopt (or do not adopt) existing technologies, as the motivation to counter surveillance may be enough for users to overcome some of the lesser usability problems found within the existing software.

\subsection{Recommendations for future research}

The variations and biases inherent in our study (Section~\ref{sec:lim}) are a consequence of the nature of retrospective research and the difficulty in recruiting from at-risk groups, and as such our research should be viewed as exploratory. While the variation may be a confounding factor, at the same time, it mimics what happens in the real world: people are trained to use technologies in different settings according to their needs and may adopt technologies for different periods of time.  In the end, this is exploratory work, but provides a richer base of information vis-\`a-vis lab studies.

Existing controlled lab studies that evaluate PGP have unveiled important usability issues with PGP and have provided a framework for future improvements in the design of secure communication.
However, the highly controlled nature of these studies  neglects to address the use of PGP in the hands of users with a threat model lending itself to motivation to adopt secure communication in the moment.
Here we have found a higher adoption rate and long-term use of PGP than one would expect from the Johnny Can't Encrypt papers, which focus on the impact of usability on learnability.
While the learnability challenges of these tools cannot be discounted, we have found that motivated users are able to overcome usability issues in existing software.

Until applications follow the practices of success stories such as the Panama Papers collaboration, and are made more appropriate for all user groups, we must figure out how to support users with tools that are readily available.
Researchers should take greater efforts to understand tools over long periods of time, in the field, to understand usability from more than the angle of learnability.  Such work can complement rigorous, controlled lab studies of convenience samples in order to form a holistic picture of the use of these secure communication technologies.

The timing of our research was during a time in which newer, easier E2EE communications technologies (namely, secure messengers) were becoming available and popular~\cite{greenberg_signal_2015}.
These applications fare better than PGP in terms of network effect, and in fact this is a bigger influence in adoption than privacy or security~\cite{abu-salma_obstacles_2017, luca_expert_2016}.  While the factors we uncover as having an impact on adoption or non-adoption of PGP (EASE, RISK, MOT-I, COMMS) are not specific to PGP and would likely play a role with other technologies, we expect that EASE may play less of a role with an easier technology.  The reduction in importance of EASE may have non-trivial impacts on the importance of RISK, MOT-I, and COMMS or on other factors.  Replicating this or a similar study with other technologies would shed further light.  This would complement the ongoing work of understanding the security needs of particular groups, such as Elliott and Brody's study of surveillance concerns of African-American New Yorkers~\cite{elliott_design_2016}, Sierra's survey of Mexican
journalists and bloggers~\cite{sierra_digital_2013}, Samarin et al.'s survey of civil society organizations~\cite{samarin_cybersecurity_2020}, as well as Citizen Lab's continued work understanding the threats faced by civil society organizations.

\subsection{Recommendations for trainers}

Whether these factors play precisely the same role for other technologies, we believe they will still have an impact on adoption.  Therefore, we recommend that trainers should take into account the attributes of trainees (e.g.\ risk, comfort with technology, motivation) in tailoring their training to make sure trainees are adequately supported to maximize tool adoption. While PGP email does pose unique technical challenges for users~\cite{mauries_dead_2017},  in addition to helping users work past the technical facets, trainers should prioritize motivating trainees about the importance of secure communication, and supporting those who are less confident in their technological skills. It is worth noting here that one's perception of their own skills can be more influential than their actual ability or experience with technology. As such, these users have the required ability but need to be reassured by trainers. Indeed, we should not ask users who require the functionality of tools with poor usability to wait on design improvements that may not come to fruition in time to benefit those users. Our study population has shown that PGP was a viable option before Signal became available and widely adopted by activist communities.

Counter to models of security practices and teachings, risk may be negatively associated with secure communication adoption for this population.
This could be because those that participate in real-world activities that are high risk simply have a higher tolerance for risk, and therefore are comfortable with taking more risks with their communication as well. This warrants further study: can this tendency be overcome? Formulae~(\ref{eq:pars-encr}) and~(\ref{eq:pars-nencr}) indicate directions for investigation: for example, can trainings reduce the technical barrier sufficiently and increase individual motivation?

\paragraph*{Acknowledgments} This material is based upon work supported by the National Science Foundation under Grant No. 1915768."  We thank our partnership with the Civil Liberties Defense Center.

\bibliographystyle{abbrv}
\bibliography{motivated}

\appendix

\section{Survey Instrument}\label{sec:instrument}
\begin{enumerate}

\item Why did you want to learn to use email encryption?  What was your reason for the training?  (text box)

\item Are you part of a community of people who use technologies for secure communication?
(yes/no)

\item Are you part of a community of people who use email encryption? (yes/no)

\item If the answer is yes to either of the previous two questions, describe how decisions are made in this community?  Please describe the decision-making process in any of the communities or groups that you belong to that use email encryption or technologies for secure communications, more generally. (text box)

\item Please rate your experience in using Thunderbird+Enigmail for email encryption (5-point Likert from strongly disagree to strongly agree):
  \begin{enumerate}
    \item I remember the basics of what Thunderbird+Enigmail is.
\item I think that I would like to use Thunderbird+Enigmail frequently.
\item I found Thunderbird+Enigmail unnecessarily complex.
\item I thought Thunderbird+Enigmail was easy to use.
\item I think that I would need the support of a technical person to be able to use Thunderbird+Enigmail.
\item I found the various aspects of Thunderbird+Enigmail were well integrated.
\item I thought there was too much inconsistency in Thunderbird+Enigmail.
\item I would imagine that most people would learn to use Thunderbird+Enigmail very quickly.
\item I found Thunderbird+Enigmail cumbersome to use.
\item I felt confident using Thunderbird+Enigmail.
\item I needed to learn a lot of things before I could get going with Thunderbird+Enigmail.
 \end{enumerate}

\item Do you continue to use Thunderbird+Enigmail?  (yes/no)

  Please answer questions 7 through 9 with respect to the Thunderbird+Enigmail system:

\item {\bf After} the training, have you done any of the following tasks?  For each, select all of the following responses that apply:
\begin{itemize}
	\item I have done this.
	\item I have not done this.
	\item I can't recall if I have done this.
	\item I don't know what this means.
\end{itemize}
\begin{enumerate}
	\item sent someone your public key
	\item received a public key from someone
	\item retrieved someone's public key from a keyserver
	\item downloaded someone's public key from a website
	\item verified the fingerprint of someone's public key
	\item verified your public key fingerprint with someone else
	\item received an encrypted or signed message with an attachment
	\item sent an encrypted or signed message with an attachment
	\item sent an encrypted message to multiple people at the same time
	\item backed up your private key
	\item created a new public/private key pair
	\item mistakenly sent an unencrypted email that you intended to be encrypted
	\item tried to send an encrypted message but were unable to
	\item sent an encrypted message that the recipient was unable to read
	\item received a public key that you were not able to use/import
	\item received a message that you couldn't decrypt
	\item received a message with an encrypted attachment that you couldn't decrypt
	\item forgot your passphrase
	\item asked someone for help with email encryption
	\item helped someone else with email encryption
\end{enumerate}

\item For any of the difficulties indicated above, please give any applicable details that you recall.  (text box)

\item Did you have any other problems with email encryption using Thunderbird+Enigmail after the training? (text box)

\item Do you use an email encryption system other than Thunderbird+Enigmail?  Which system, and why do you use it? Please list and discuss all systems you use and why.  (text box)

\item Which email encryption system do you use most frequently use? (text box)

  [Questions 7-9 asked with respect to the system that used most frequently]

\item If you no longer use Thunderbird+Enigmail, why?  (text box)

\item If you no longer use email encryption, why?  (text box)

\item For the following questions, select all that apply (never/in the past/sometimes/often/always):
  \begin{enumerate}
  \item Do you encrypt emails even if their contents aren't considered personal/sensitive?
  \item Do you use email encryption for personal email?
  \item Do you use email encryption for activist/organizing/volunteer work?
  \item Do you use email encryption for paid work?
  \end{enumerate}

\item Which of the following ways do you regularly access the email account(s) that you use for sending and receiving encrypted emails, e.g. to read or send unencrypted email from the same account (select all that apply):
\begin{itemize}
\item Thunderbird
\item Outlook
\item Apple Mail
\item iPhone/iPad
\item Android device
\item web interface (e.g. via gmail.com or mail.riseup.net)
\item other: (text box)
\end{itemize}
\item Imagine you are starting to use a new program (or app) on your laptop or desktop.  For each of the following, indicate whether your level of confidence in being able to use the new program. (``I couldn't do it'' plus a 10 point Likert from ``Not at all confident'' through ``Moderately confident'' to ``Totally confident'', as a table)

  I could use this new program ...
  \begin{enumerate}
  \item if there was someone giving me step-by-step instructions.
  \item if there was no one around to tell me what to do as I go.
  \item if I had never used a program like it before.
  \item if I had only the program manuals for reference.
  \item if I had seen someone else using it before trying it myself.
  \item if I could call someone for help if I got stuck.
  \item if someone else had helped me get started.
  \item if I had a lot of time to use the program.
  \item if I just had access to built-in help in the program.
  \item if someone showed me how to do it first.
  \item if I had used similar programs before this one.
  \end{enumerate}
\item What is your gender identity?  (text box)

\item What is your ethnic identity?  (text box)

\item How would you describe your political orientation?  (text box)

\item What is your age bracket (select one): \\
\hfill $\le$ 19 \hfill 20-29 \hfill 30-39 \hfill 40-49 \hfill 50-59 \hfill 60-69 \hfill 70+ \hfill

\end{enumerate}
\newpage
\begin{figure*}[h]
  \centering
  \vspace*{-.5in}
  \hspace*{-.5in}
  \includegraphics[trim=50 30 50 30,clip, width=8in, angle = 90]{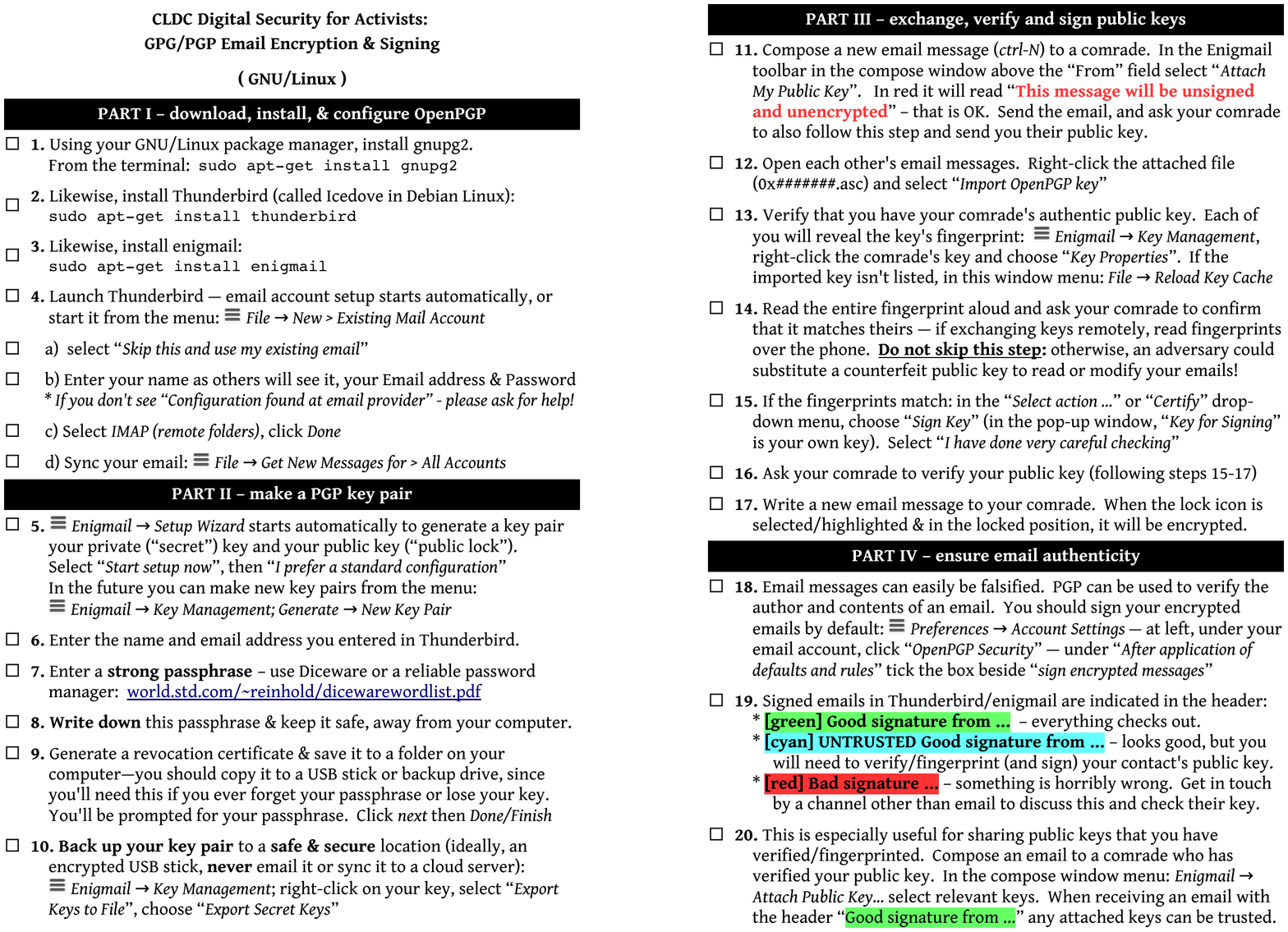}
  \caption{PGP worksheet for Linux.  Similar worksheets were provided for Windows and Mac.}
\end{figure*}

\end{document}